\documentstyle[prl,aps,twocolumn,floats]{revtex}
\begin{document}
%\draft
\title{Adiabatic geometric phases and response functions}
\author{ Sudhir R. Jain and Arun K. Pati }
\address{Theoretical Physics Division, Central Complex, Fifth Floor}
\address{Bhabha Atomic Research Centre,
Mumbai 400 085, India}
\maketitle

\begin{abstract}
Treating  a  many-body Fermi system in terms of a single particle
in  a   deforming  mean-field,  we  relate  adiabatic
geometric phase to susceptibility for the non-cyclic case, and to
its  derivative  for the cyclic case. Employing the semiclassical
expression  of  the  susceptibility,  the expression for geometric phase for
chaotic   quantum  system  immediately  follows.  Exploiting  the
well-known association of the absorptive part  of  susceptibility
with   dissipation,  our  relations  may  provide  a  quantum
mechanical origin of the damping of collective excitations in
Fermi systems.
\end{abstract}
\vskip 2.5 truecm

\vskip 0.5 truecm
\noindent
Email : krsrini@magnum.barc.ernet.in
\newpage

Chaotic adiabatic systems \cite{swiatecki} serve as useful models
for  complex systems in the mean-field approximation. These idealized
models  have  been  employed  to  understand   very   interesting
phenomena  in  nuclear \cite{hill} and plasma physics \cite{ott}.
In particular, an important problem of  many-body  theory  is  to
relate  the slow collective excitations to faster single-particle
motions. Adiabatic approximation  leads  us  to  linear  response
theory   on   one   hand   where   dynamical  susceptibility  (or
polarization propagator) \cite{fetter} is central, and  geometric
phases on the other. A relation between these seemingly disparate
quantities is being sought for in this Letter.

Phase  factors  of geometric origin were discovered for adiabatic
quantum systems \cite{berry,mead} and  have  been  generalised  to
non-adibatic  situations  \cite{ya}.  The geometry of the Hilbert
space plays a key role  in  understanding geometric  phases
\cite{gm},   a  recent  illustration being for the Josephson  junction
\cite{ja}. Adiabatic geometric phase led to an  understanding  of
several  phenomena  in  physics  \cite{shapere}  like  fractional
statistics  in  two-dimensional  statistical  mechanics,  integer
quantum  Hall effect, anomalies in field theory, the Magnus force
in the context of superfluidity  \cite{thouless} and so on.
Recently, it has been shown that the viscosity  of  quantum  Hall
fluid  in  two  dimensions  at  zero  temperature  is  related  to
adiabatic curvature (whose flux gives the  phase)  \cite{av}.  An
interesting  aspect of many-body physics related to these
advancements is in the Born-Oppenheimer approximation where reaction
forces are shown to be of geometric origin  \cite{robbins}.  Owing
to  an  intimate tangle of collective and single-particle
effects, it becomes useful to treat the total system as one where
a slow subsystem is coupled to a faster  one.  In  this  setting,
first-order  corrections  to  the  Born-Oppenheimer approximation
leads to geometric magnetism and deterministic friction in a fully
classical treatment, whereas a half-classical treatment gives
rise to geometric magnetism only.

When  a  particle (a nucleon, say) moves inside an enclosure whose
boundary is adiabatically vibrating in time, the wavefunction can
acquire a geometric phase over a cycle of vibration.  This  model
is   an  idealization  of  a  single  particle  in  a  mean-field
represented by the enclosure, and has been  an established  paradigm
for  numerous  studies  in  the  past  \cite{swiatecki}. We would
particularly like  to  emphasize  here  the  importance  of  such
studies  in  enhancing the understanding of damping of collective
excitations in nuclear physics. Concepts like  time  correlations
and  susceptibility  are fundamental to any discussion of quantum
statistical mechanics as they lead to  an  understanding  of  the
response   of   many-body   system  \cite{fetter}  and  the transport
coefficients \cite{kubo}. It is  well-known  that  the  imaginary
part  of susceptibility is related to dissipation, thus it may be
quite interesting to explore its possible relationship with
a quantity like geometric phase. To work out such a relation,
we  begin  with  a discussion of geometric phase and identify the
appropriate   quantities related to
susceptibility.

Let us first consider a Hamiltonian parametrised by ${\bf R}$ which describes a
single particle in an effective mean-field where the shape of the field
is vibrating adiabatically in time. It is well-known \cite{berry} that when
the parameters evolve along a cyclic path, ${\cal C}$, the
instantaneous eigenfunction of the system, $ |n({\bf R})\rangle $ corresponding
to the eigenvalue $E_n({\bf R})$, acquires a
geometric phase given by
\begin{equation}
\gamma  _n({\cal C}) = \oint_{\cal C} i\langle n({\bf R}|\nabla_{\bf R} n({\bf R})\rangle .d{\bf
R} = - \frac{1}{\hbar } \int_{\cal S} {\bf V}_n.d{\bf S},
\end{equation}
where ${\cal S}$ is the surface enclosed by ${\cal C}$ in the parameter space, and
${\bf V}_n$ is the "field-strength" (adiabatic curvature) given by a familiar expression
involving a sum of weighted wedge product between two appropriate
matrix elements :
\begin{equation}
{\bf V}_n = -i\hbar \sum_{m(\neq n)}\frac{\langle  n|\nabla _{\bf R}H|m\rangle
\wedge \langle  n|\nabla _{\bf R}H|m\rangle }{(E_n-E_m)^2}.
\end{equation}
A suitable form of ${\bf V}_n$ for the sequel is \cite{jr}
\begin{equation}
{\bf V}_n = \frac{i}{2\hbar } \lim_{\epsilon \rightarrow 0} \int_{0}^{\infty}
 dt e^{-\epsilon t} t \langle  n|\left[(\nabla _{\bf R}H)_t,\wedge (\nabla _{\bf R}H)
 \right]|n\rangle
\end{equation}
where $(\nabla _{\bf R}H)_t$ denotes the Heisenberg-evolved operator.
Note that, the state $|n({\bf R})\rangle $ appearing in  (3) corresponds to a single-particle
eigenket in an effective mean-field. This state is clearly related to the
original many-body Fermi system for which the imaginary part of the
dynamical susceptibility is \cite{kubo}
\begin{eqnarray}
\chi ^"(t) &=& \frac{1}{2\hbar }\langle  \Phi_0|[\hat{\cal A}(t),\hat{\cal B}(0)]|\Phi_0 \rangle
\nonumber \\
&=& \int \frac{d\omega }{2\pi }e^{-i\omega t}\tilde{\chi }^"(\omega )
\end{eqnarray}
where $|\Phi _0\rangle $ is the pure ground state of the many-particle system
with Fermi energy, $E_F$.
If one-body operators, $\hat{H}$, $\hat{A}$, and $\hat{B}$,
are  used  to construct many-body operators by a direct sum so as
to get
$\hat{\cal    H}$,   $\hat{\cal   A}$,   and    $\hat{\cal   B}$,
respectively,
and
$\hat{\cal H}|\Phi _l\rangle ={\cal E}_l|\Phi _l\rangle $, then we \cite{pg-sj}
have
\begin{equation}
\langle \Phi _0|\hat{\cal A}|\Phi _l\rangle  = \langle m({\bf R})|\hat{A}|n({\bf R})\rangle .
\end{equation}

On reducing the many-body system at $T = 0 K$, (where the Fermi-Dirac distribution
is a Heaviside step function), to one-body system, we can express \cite{pg-sj}
\begin{equation}
\tilde{\chi }^"_{AB}(\omega  ) = -\frac{\omega }{2}\int  dt
e^{i\omega t}~\mbox{tr~}\delta (E_F-H)[\hat{A}(t),\hat{B}(0)].
\end{equation}
This can also be written semiclassically as \cite{pg-sj}
\begin{eqnarray}
\tilde{\chi }^"_{AB}(\omega  ) &=& -\frac{\omega }{2}\int dt \int
\frac{d^fxd^fp}{(2\pi \hbar )^f}\delta (E-H)A_W({\bf x},{\bf p})  \nonumber \\
&~&\left[
e^{(i\omega - \hat{\cal L}_{cl})t}B_W\right]({\bf x},{\bf p}) + {\cal O}(\hbar ^{-f+1})
\nonumber \\
&-& \frac{\omega }{2\pi \hbar }\delta (E_F-H)\sum_{p,r}
\frac{\cos \left(\frac{r}{\hbar}S_p-r\frac{\pi}{2}\nu _p \right)}
{|\mbox{det}({\bf m}_p^r-{\bf I})|^{1/2}}.     \nonumber \\
&.&\int dt e^{i\omega t}\oint_p d\tau A_W(\tau )B_W(\tau + t) + {\cal O}(\hbar ^0)
\end{eqnarray}
where the subscript $'W'$ refers to the Weyl symbol of the operator, $f$ denotes
the degrees of freedom and $\hat{\cal L}_{cl}$ is the Liouvillian operator.
The last term corresponds to the periodic orbit expansion where $S_p$,
$\nu _p$, and ${\bf m}_p$ correspond to action, Maslov index, and the monodromy
matrix for the periodic orbit, $p$, and $r$ denotes the repetitions. We have used
the Gutzwiller trace formula \cite{gutzwiller} for the case where the single-particle
motion is chaotic. The above semiclassical expression is valid for Hamiltonian
operators which are quadratic in momentum, $\hat{p}$, with an additive
position-dependent
term. The expression (6), however, is general.
For the case where the system has partially broken
symmetry, the results have been recently generalized \cite{sj}.

The label '$n$' in (3) corresponds to single-particle states and is related to
$|\Phi _0\rangle $ because the many-body matrix elements can be written in terms of
one-body matrix elements for the case when all the constituents are
taken as non-interacting. In many-body physics, this gives the zero-order
response whereupon the interaction can be included in a Vlasov description
in an iterative way \cite{brink}. For relating the response function to the
geometric phase, the operators $\hat{\cal A}$ and
$\hat{\cal B}$ in our discussion are to be identified with $\nabla _X\hat{\cal H}$ and
$\nabla _Y\hat{\cal H}$ for ${\bf R}=(X, Y, Z)$.

The matrix element in (3) can be written as a many-body matrix element
using (5) by composing $\hat{A}$ and $\hat{B}$ so that we get the operator,
${\cal C}(t)=[{\cal A}(t),{\cal B}(0)]-[{\cal B}(t),{\cal A}(0)]$,
which s related to a difference $\chi "_{AB}(t)-\chi "_{BA}(t)=\chi "_C(t)$.
Thus, we can re-write ${\bf V}_n$ as
\begin{eqnarray}
{\bf V}_n &=& \frac{i}{2\hbar } \lim_{\epsilon \rightarrow 0} \int_{0}^{\infty }
dt e^{-\epsilon t}~t~\langle \Phi _0|{\cal C}|\Phi _0 \rangle \nonumber \\
&=& \int_{0}^{\infty } dt~t\chi "_C(t)
= -\frac{\partial \tilde{\chi }"(\omega)}{\partial \omega }\vline_{\omega = 0}.
\end{eqnarray}
We now arrive at our first relation for the case of cyclic evolution :
\begin {equation}
\gamma _n({\cal C}) = \int_{\cal S} d{\bf S}.
\frac{\partial \tilde{\chi }"_{C}(\omega ;{\bf R})}{\partial \omega }
\vline_{\omega =0}.
\end{equation}
Since $\tilde{\chi }"_{C}(\omega )$ is related to energy dissipation,
this relation connects geometric phase to dissipation.

Exploiting (7), we get
\begin{eqnarray}
&&\frac{\partial \tilde{\chi }^"_{AB}(\omega ;{\bf R})}{\partial \omega }
\vline _{\omega =0} = -\frac{1}{2}\int dt \int
\frac{d^fxd^fp}{(2\pi  \hbar  )^f}
\delta  (E_F-H)\nonumber \\ && [A_W\left(
e^{-{\cal L}_{cl}t}B_W\right) - B_W\left( e^{-{\cal L}_{cl}t}
A_W\right)] \nonumber \\
&&- \frac{1}{2\pi \hbar }\lim_{\omega \rightarrow 0} \delta (E_F-H)\sum_{p,r}
\frac{\cos \left(\frac{r}{\hbar}S_p-r\frac{\pi}{2}\nu _p \right)}
{|\mbox{det}({\bf m}_p^r-{\bf I})|^{1/2}} \nonumber \\
&&\int dt e^{i\omega t}\oint_p d\tau
\left[A_W(\tau )B_W(\tau + t)
- B_W(\tau )A_W(\tau + t)\right]
\end{eqnarray}
which   entails   the   semiclassical  expression  for  adiabatic
geometric  phase  for  chaotic systems, using  (9). The first term comes from
the  classical  time  correlation  function  which is expected to decay
exponentially  as the dynamics is chaotic. This decay is governed
by the largest Liapunov exponent of the classical system. The  second  term
is the semiclassical correction in terms of periodic orbits which
can be termed as a periodic orbit two-form.   It is important to note that (9)
is valid for general Hamiltonians whereas (10) is restricted to the operators like
$\hat{p}^2 + V(\hat{q})$ only.

The time-dependent deformation of the mean-field can, in general,
be  non-cyclic,  particularly  due  to  the  fact  that   several
harmonics  (possibly  incommensurate)  may  be  involved. In this
general case, the above expression  does  not  hold  and  we  now
proceed to quantify the general relation.

When  a  quantal  system  does  not follow a cyclic evolution the
geometric phase not only depends on the  path  of  the  evolution
curve  but  also  on  the  end  points.  It  has  been  shown in
\cite{pati} that the wavefunction of the  system  can  acquire  a
geometric  phase  during  its  time  evolution,  irrespective  of
adiabatic and cyclic condition, which is given by
\begin {equation}
\gamma(\Gamma) = \int i\langle \chi(t)|{\dot \chi}(t)\rangle dt,
\end{equation}
where $|\chi(t)\rangle $ is a modified state  vector  defined  from  the
actual  state   vector  $|\Psi(t)\rangle $ of the system as $|\chi(t)\rangle  =
{\langle \Psi(t)|\Psi(0)\rangle   \over  |\langle \Psi(t)|\Psi(0)\rangle |}|\Psi(t)\rangle $
and an overdot denotes the time-derivative. From
this general expression, the adiabatic, open-path geometric phase can be written
in the following way,
\begin {equation}
\gamma  _n(\Gamma) = \int_{\Gamma } i\langle \chi_n({\bf R})|\nabla \chi_n({\bf
R})\rangle .d{\bf R} = \int_{\Gamma } \Omega_n({\bf R}).d{\bf R},
\end{equation}
where $\Omega_n({\bf R})$ is  a  generalised  adiabatic  vector
potential   (connection  one-form)  whose  line  integral  gives  the
geometric phase. This non-cyclic adiabatic geometric phase can be
expressed  as  a  line  integral  of  the difference of two
vector potentials \cite{akp},
\begin {equation}
\gamma  _n(\Gamma) = \int_{\Gamma } ({\bf A}_n({\bf R}) -{\bf
P}_n({\bf R})).d{\bf R},
\end{equation}
where ${\bf A}_n({\bf R}) =  i\langle n({\bf  R})|\nabla  n({\bf
R})\rangle $  is  the usual Berry potential whose curl gives the adiabatic
curvature ${\bf V}_n$ and
${\bf P}_n({\bf R}) = -\Im ({\langle n({\bf R}(0))|\nabla  n({\bf
R})\rangle  \over \langle n({\bf R}(0))|n({\bf R})\rangle })$ is an extra potential
that  takes  care of the contributions from the end points of the
evolution  path.  The  non-cyclic  geometric   phase   is   gauge
invariant because under a local gauge transformation,
${\bf  A}_n$  and ${\bf P}_n$ transform in the same way. Also
in the special case of cyclic evolutions of  parameters, (12)
goes over to the well known expression, (1).

\par
For our purpose, we simplify the generalised vector potential,
$\Omega_n({\bf R})$, as
\begin {eqnarray}
\Omega_n({\bf R}) &=& {\bf A}_n({\bf R}) -{\bf
P}_n({\bf R})\nonumber \\ &=&
\Im \sum_{m(\neq n)} {\langle n({\bf R}(0))|m({\bf R})\rangle  \over \langle n({\bf R}(0))|n({\bf
R})\rangle } {\langle m({\bf R})|\nabla H|n({\bf R})\rangle  \over (E_n - E_m)}.
\end{eqnarray}

\par
Using an integral representation of $(E_n - E_m)^{-1}$,  we can write
$\Omega_n({\bf R})$ as follows
\begin {eqnarray}
\Omega_n({\bf R}) &=& {1 \over \hbar}
\sum_{m(\neq n)}
\lim_{\epsilon \rightarrow 0} \int_{0}^{\infty }
dt e^{-\epsilon t} \Im \bigg[ i e^{-i(E_n -E_m)t/\hbar} \nonumber \\
&~&\frac{\langle n({\bf R})|P_n({\bf R}(0))P_m({\bf R}) \nabla H|n({\bf R})\rangle }{
|\langle n({\bf R}(0))|n({\bf R})\rangle |^2} \bigg]
\end{eqnarray}
where $P_n({\bf R}(0)) = |n({\bf R}(0))\rangle \langle n({\bf R}(0))|$ and  $P_m({\bf R}) = |m({\bf R})\rangle \langle m({\bf R})|$
are  the  projection  operators  corresponding  to  $n$th  and  $m$th
eigenstates. Now define a quantum  correlation  function  between
the operator, $A$, and the Heisenberg evolved operator, $B_t$, through
\begin {equation}
Q_{AB}(t) = {1 \over 2}\langle n|(AB_t + B_tA)|n\rangle  -  \langle n|A|n\rangle \langle n|B_t|n\rangle
\end{equation}
and identify the Hermitian operators $A = P_n({\bf R}(0))$ and $B =
\nabla H$. With this, the  generalised  vector  potential  can  be
written in terms of a quantum correlation function:
\begin {equation}
\Omega_n({\bf R}) = {1 \over \hbar}
\lim_{\epsilon \rightarrow 0} \int_{0}^{\infty }
dt e^{-\epsilon t} {Q_{AB}(t) \over |\langle n({\bf R}(0))|n({\bf R})\rangle |^2}
\end{equation}
Let  us  note  that  $Q_{AB}(t)$   defines fluctuations in the
symmetrised  correlation  of  two  operators.  Recalling   the
relationship between the Fourier transform of $Q_{AB}(t)$ and the
susceptibility $\tilde{\chi }"(\omega )$ \cite{forster}  at  zero
temperature, we find
\begin {equation}
\tilde{Q}_{AB}(\omega) = \hbar \tilde{\chi }"(\omega ),
\end{equation}
an  instance of the fluctuation-dissipation theorem of the second
kind  \cite{kth}.  Following  the  physical  arguments  used   to
express
single  particle  expectation  values  in terms of statistical
quantities as in the cyclic case, we get
\begin {eqnarray}
\tilde{\chi }"(0;{\bf R}) &=& {1 \over \hbar}
\lim_{\epsilon \rightarrow 0} \int_{0}^{\infty }
dt e^{-\epsilon t} Q_{AB}'(t) \nonumber \\
&=& \Omega_n({\bf R}),
\end{eqnarray}
where $Q_{AB}'(t) = {Q_{AB}(t)\over |\langle n({\bf R}(0))|n({\bf R})\rangle |^2}$ is the scaled
quantum  correlation  function. Here, the scale  factor is the survival
probability  $P(t)  =  |\langle n({\bf  R}(0))|n({\bf  R})\rangle |^2$ which is
related to
the  two-time  correlation  function  at  zero
temperature for a system in  canonical  equilibrium \cite{jgv}.  This  result
seems  to be a fundamental one as the generalised adiabatic vector
potential  is  exactly  equal  to  the  imaginary  part  of   the
susceptibility   at  zero  frequency.  Therefore,  the  adiabatic
geometric phase for a non-cyclic variation of external parameters along
a path $\Gamma$ is
given by
\begin {equation}
\gamma_n(\Gamma)  =  \int_{\Gamma}  \tilde{\chi }"(0;\bf R).d{\bf
R},
\end{equation}
which relates  a  bulk  property  of  the  many-body  Fermionic,
chaotic  quantum  system  to  a  physical quantity assigned to a
single particle.

In (9) and (20), note that the frequency-dependent response functions
are different. However, they are related as follows :
for a cyclic evolution along a path, ${\cal C}$, applying Stokes'
theorem, (20) reduces to a surface integral of
curl$\tilde{\chi }"(0;{\bf R})$ over ${\cal S}$. Using (9), it follows that
\begin{equation}
\nabla _{\bf R} \times \tilde{\chi }"(0;{\bf R}) =
\frac{\partial \tilde{\chi }"_C(\omega ;{\bf R})}{\partial \omega }\vline_{\omega = 0}.
\end{equation}

The  above  results  have clear relevance to finite Fermi systems
like metallic clusters  and  nuclei.  The  historical  collective
picture  of  nucleus  was  Bohr's  liquid  drop  which  has  been
successfully used ever since. Due to  the  numerical  simulations
\cite{swiatecki}  and  the  results  (eqs.(9)  and (20)) obtained
above,  it  is  clear  that  the  nuclear  fluid  has,  in  fact,
characteristics  of  a  gel which is an elastic or a viscoelastic
liquid.  The  connecting  relation  (21)   between   cyclic   and
non-cyclic  case  is  necessary as a consistency requirement. The
applicability of our results to nuclei is due to  the  compelling
evidence,  originating  from shell structure of binding energies,
testifying that the mean free path of nucleons is larger than the
nuclear dimensions \cite{bracco}. Hence it is natural to look for
the source  of  viscosity  in  the  nuclear  fluid  in  terms  of
single-particle  features where the effective mean-field leads to
chaotic dynamics.

\par
It is well known that viscosity or friction in a quantal  context
can  appear  in  thermodynamic  limit. Since we have
consistently discussed many-body Fermi systems in equilibrium,
the  seemingly  contradictory  conditions  of  finiteness  of the
system and ``continuity'' of the energy spectrum are  met  with.  The
finite-size of the system explicitly manifests itself in terms of
the sum over periodic orbits. Thus, yet another way to  look  at  our
relations  (9),  (20) is that these relations are off-springs of
finite-size  and  spectral  continuity-  a  stylized  version  of
thermodynamic limit  where viscosity and geometric phase can be
discussed together.

\par
The notion of viscoelasticity in the context of  nuclear  physics
has been pointed out in Ref. [1]. If we accept that the model  for
nucleus  is  a  gel  model  and  apply linear viscoelastic theory
\cite{doi} where stress relaxation function is connected  to  the
time-correlation  function,  we  can  see  that  the generalised
vector potential (connection one-form) is directly related to
viscosity. This is due to the fact that the time integral of  the
relaxation  function  is  identified  with  viscosity  in  linear
viscoelastic theory.

\par
Every time some relationships with possible fundamental relevance
are  given,  the  question  about  their  generality  arises. The
relation (9) and (20), so-to-say ($\gamma_n$-$\chi$)  relations,
are  restricted  by the limits $T \rightarrow 0$.
For any non-zero $T$, the semiclassical  result
for susceptibility is already known \cite{pg-sj}. We believe that
the  ($\gamma_n$-$\chi$)  relation will be given as above where
$\tilde {\chi}''$ will be evaluated at some non-zero value of $\omega $
consistent with the temperature. Hence, as long as  the  dynamics
is  chaotic,  these  relations  hold.  This  takes  us  to a deep
interplay of quantal and statistical  features  embodied  in  the
subject  of quantum chaos. It is interesting to mention here that
recent works have led to a new treatment of  quantum  statistical
mechanics in terms of Berry conjecture \cite{ms,srj-da}. Further,
it  has been shown \cite{ms1} that thermal fluctuations for large
number of degrees of freedom coincide  with  quantum  fluctuations
for  chaotic  quantum systems in the semiclassical limit when the
freedoms are  smaller in number.  One  can  immediately  rationalise  above
discussion in the light of these contemporary works. In the above
discussion, when we pick up a single-particle state, it ought to
be one that can be written as a random superposition of plane waves
in the sense of \cite{ms}. This is possible for a system with
chaotic classical dynamics. It is these states which can be combined
into a Slater determinant leading to the Fermi-Dirac distribution.

In conclusion we have established  a  fundamental relation between
adiabatic geometric phase and response function for cyclic as
well as non-cyclic evolutions of chaotic  quantum  systems  where
chaos   appears   as  a  result  of  the  mean-field  seen  by  a
single-particle due to other particles. Since the imaginary  part
of the susceptibility is related to energy dissipation, our result
provides  a  quantum  mechanical  way to interpret and understand
decay of collective excitations of Fermi systems. The results  of
this  Letter can be generalised for the case when the dynamics is
not fully chaotic using the results of [17].

\end{document}